\documentclass{iaus}
\usepackage{graphicx}

\title[Disk Inhomogeneities and the Resultant Planet Traps] %% give here short title %%
{Disk Inhomogeneities and the Origins of Planetary System Architectures and Observational Properties}

\author[Yasuhiro Hasegawa \& Ralph E. Pudritz]   %% give here short author list %%
{Yasuhiro Hasegawa$^{1}$
 \and Ralph E. Pudritz$^{2,3}$}

\affiliation{$^1$EACOA fellow, Institute of Astronomy and Astrophysics, \\ 
   Academia Sinica (ASIAA), Taipei 10641, Taiwan \\ email: {\tt yasu@asiaa.sinica.edu.tw} 
   \\[\affilskip]
$^2$Department of Physics and Astronomy, McMaster University, \\
    Hamilton, ON L8S 4M1, Canada 
    \\[\affilskip]
$^3$Origins Institute, McMaster University, Hamilton, ON L8S 4M1, Canada
\\ email: {\tt pudritz@physics.mcmaster.ca}
}

\pubyear{2013}
\volume{xxx}  %% insert here IAU Symposium No.
\pagerange{1--4}
\jname{Exploring the Formation and Evolution of Planetary Systems}
\editors{B. Metthews, \& J. Graham, eds.}
\begin{document}

\maketitle

\begin{abstract}
Recent high-resolution observations show that protoplanetary disks have various kinds of structural properties 
or ÒinhomogeneitiesÓ.   These are the consequence of a mixture of a number of physical and 
chemical processes taking place in the disks. Here, we discuss the results of our comprehensive  investigations 
on how disk inhomogeneities affect planetary migration. We demonstrate that disk inhomogeneities give rise to planet traps - specific sites in 
protoplanetary disks at which rapid type I migration is halted. We show that up to three types of traps (heat transitions, ice lines and dead zones) can exist 
in a single disk, and that they move differently as the disk accretion rate decreases with time. We also demonstrate that the position of planet traps strongly depends on stellar masses and disk accretion rates. This indicates that host stars establish preferred (initial) scales of 
their planetary systems.  Finally, we discuss the possible observational signatures of disk inhomogeneities. 

\keywords{accretion, accretion disks, radiative transfer, turbulence, (stars:) planetary systems: formation, (stars:) planetary systems: protoplanetary disks.}
\end{abstract}

\firstsection % if your document starts with a section,
\section{Introduction}

The advent of "next generation" telescopes has revealed that protoplanetary disks are rich in their structures  (\cite[Tamura 2009]{Tamura09}). 
For instance, SMA observations have recently inferred that there is a jump in the column density of $^{13}$CO at CO-ice lines (\cite[Qi et al 2011]{Qi11}). 
The evidence suggests that the jump occurs due to the condensation of CO onto dust grains at the CO-ice line, where the disk temperature declines to about 160 K. 

One of the most serious problems in theories of planet formation is planetary migration (\cite[Kley \& Nelson 2012]{KleyNelson12}). 
It arises from tidal, resonant interactions between gaseous disks and protoplants that form in the disks. The most advanced studies show that 
the timescale of type I migration, that is applicable for low-mass planets such as terrestrial planets and cores of gas giants, is very rapid 
($\sim 10^5$ years for planets with $\sim 1 M_{\oplus}$ at $\sim 1$ AU), and that its direction is highly coupled with disk properties 
such as the surface density and the temperature of disks.  Here, we report on our investigations as to how disk structures or "disk inhomogeneities" slow 
planetary migration and  explore their consequences for planetary system architectures.  We also discuss how important
the inhomogeneities are by bridging between the observations of exoplanets and protoplanetary disks. 

\begin{figure}[b]
% \vspace*{-2.0 cm}
\begin{center}
 \includegraphics[width=6.0cm]{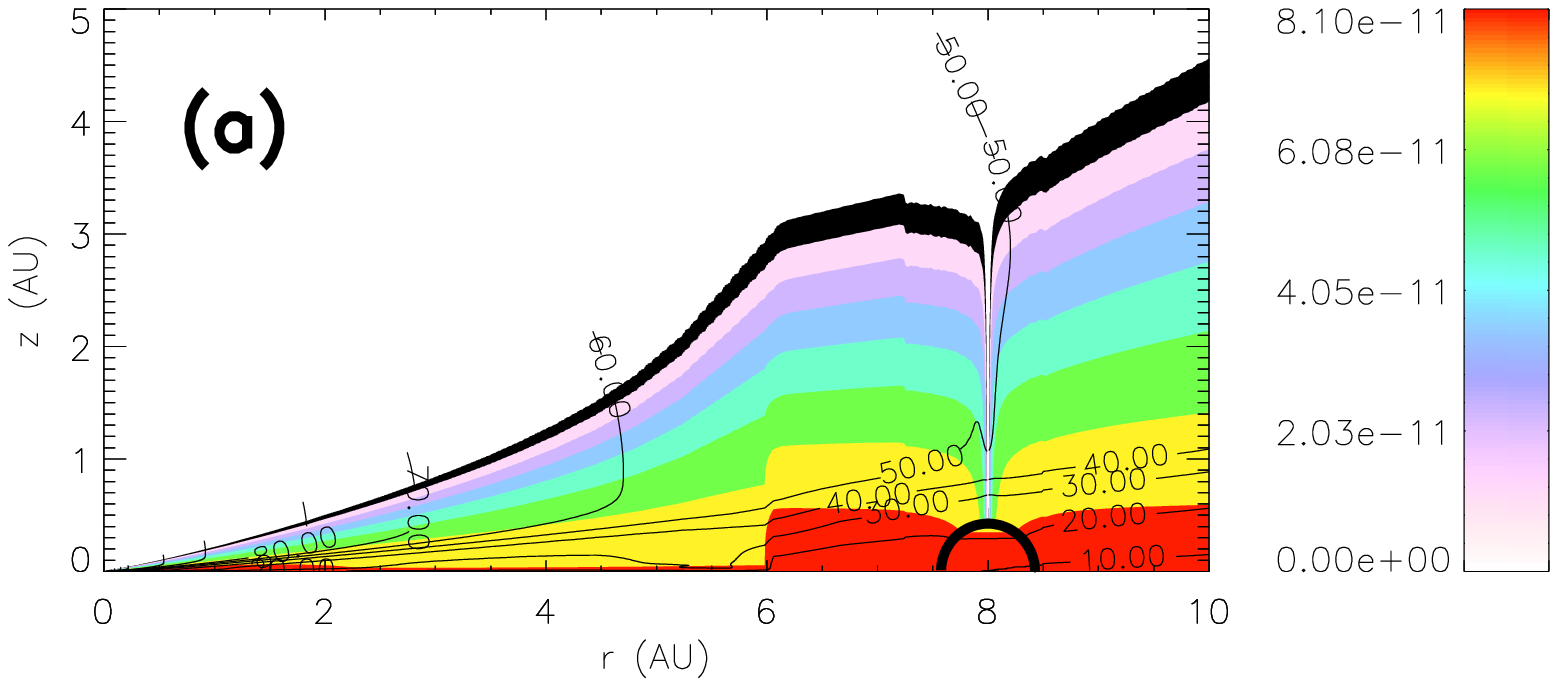}
 \includegraphics[width=6.0cm]{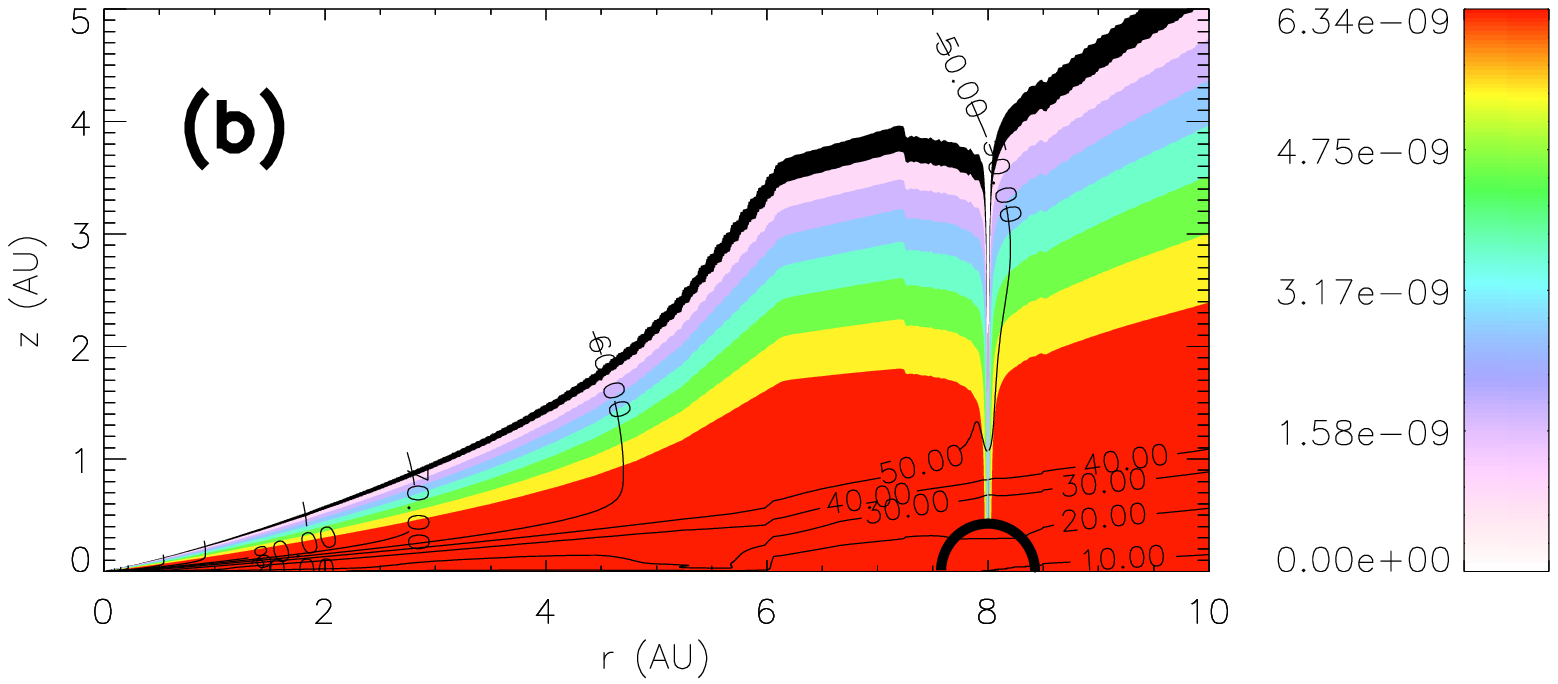}
 \includegraphics[width=6.0cm]{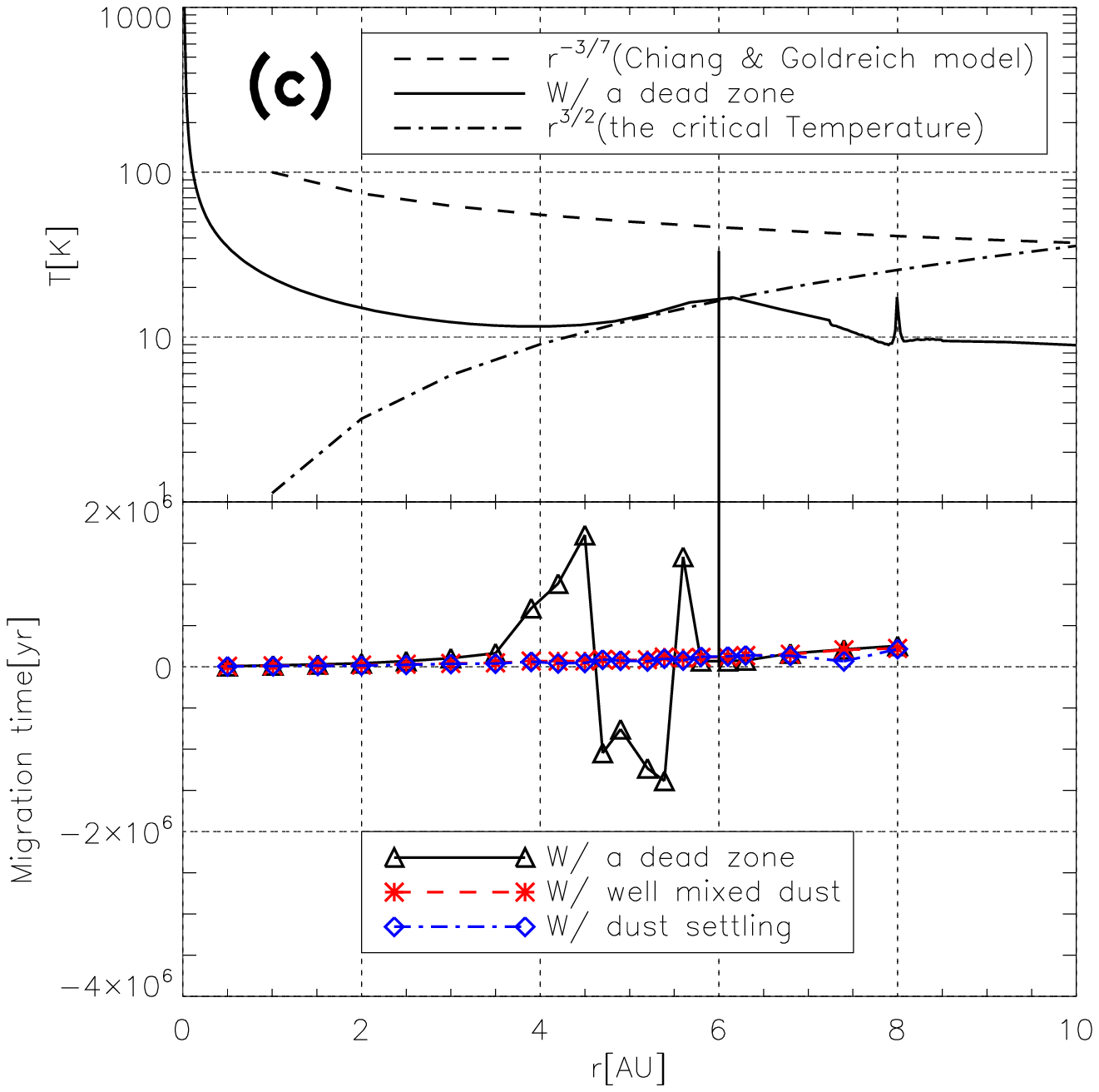}
 \includegraphics[width=6.0cm]{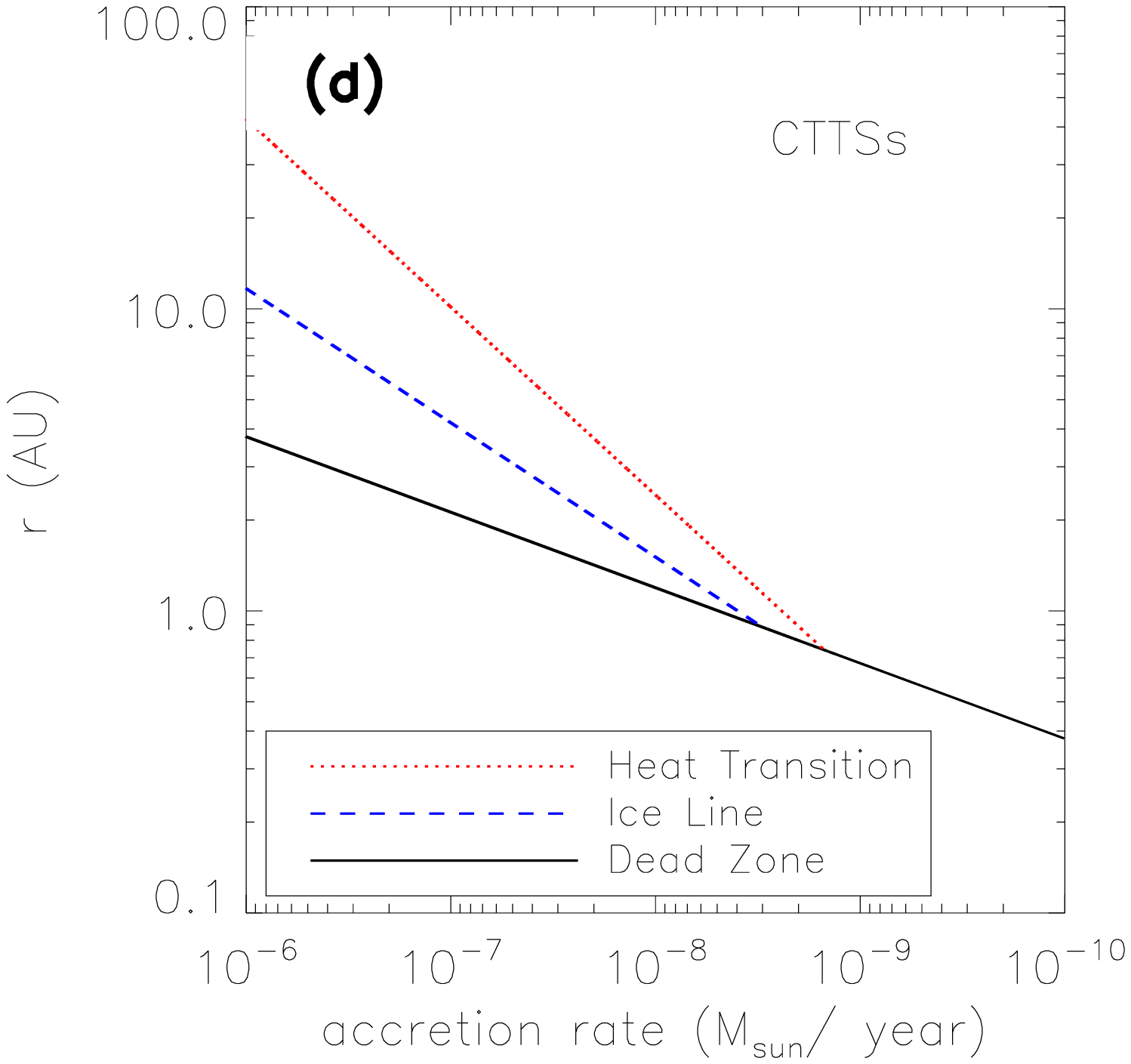}
% \vspace*{-1.0 cm}
 \caption{Radiatively heated protoplanetary disks with dead zones (in (a) and (b); adapted from \cite[Hasegawa \& Pudritz 2010a]{HasegawaPudritz10a}). 
The dust  and gas density distribution, that is denoted by colors in g cm$^{-3}$, is shown with the temperature contours in Kelvin, 
on panel a) and b) respectively. Planetary migration in radiatively heated disks with dead zones which is 6 AU in size (in (c); adapted from 
\cite[Hasegawa \& Pudritz 2010a]{HasegawaPudritz10a}). In c) (top), the temperature profile of the mid-plane region as a function 
of the distance from the central star. In c) (bottom), the timescale of type I migration as a function of position of a planet from the central star. Position of 
planet traps (in (d); adapted from \cite[Hasegawa \& Pudritz 2011]{HasegawaPudritz11}). The position of planet traps as a function of disk accretion rate for 
classical T Tauri stars (CTTSs).
}
   \label{fig1}
\end{center}
\end{figure}

\section{Disk Inhomogeneities and the Resultant Planet Traps}

We first discuss dead zones,  which are predicted to be present in the inner region of disks (\cite[Gammie 1996]{Gammie96}). 
In general, protoplanetary disks are ionized by X-rays from the central stars and cosmic rays. In the inner region of disks, however, 
the ionization is not so efficient due to the high column density there.   Magnetorotational instabilities (MRIs) will be largely suppressed there, 
because MRIs require a good coupling between charged fluids in the gas disks and magnetic fields threading the disks. 
As a result, protoplanetary disks are characterized mainly by two zones: the inner, high density region is the so-called MRI "dead" zone and is weakly turbulent 
whereas the outer, low density region is a MRI active zone and is highly turbulent.

We examine the effects of dead zones on the thermal structure of disks and investigate how planetary migration is affected by the presence of dead zones. 
Figure \ref{fig1} (panels (a) and (b)) shows the results of the density and thermal structures of disks with dead zones 
(\cite[Hasegawa \& Pudritz 2010a]{HasegawaPudritz10a}). The dust distribution (panel (a)) shows that a dusty "wall" is present 
at the outer boundary of a dead zone that is 6 AU in size. This arises because dust settling in the dead zone is significantly enhanced 
due to the low value of turbulence whereas dust in the active zone is still mixed well with the gas. 
Adopting the dust distribution, we computed the disk temperature. In order to make our calculations reliable, 
we performed Monte Carlo based, radiative transfer simulations, assuming the main heat source is stellar irradiation  
(also see \cite[Hasegawa \& Pudritz 2010b]{HasegawaPudritz10b}). Once we obtain the temperature structure of disks, 
we re-calculate the gas distribution using the computed disk temperature, assuming the vertical hydrostatic equilibrium. 
It is interesting that wall-like structures do not appear in the gas distribution (see Panel (b)).

We investigate how the dusty wall affects the thermal structure of disks. In order to proceed, 
we plotted the temperature of the midplane as a function of the distance from the central star. 
Figure \ref{fig1} (panel (c); top) shows that the disk temperature has a {\it positive} gradient in the dead zone. 
This is the major finding in \cite[Hasegawa \& Pudritz (2010b)]{HasegawaPudritz10b} and can be understood as what follows: 
in general, the central stars irradiate the surface layer of disks directly, so that the midplane region of disks is {\it indirectly} heated by such surface layers. 
When a disk has a dead zone, a dusty wall is left at the outer boundary of the dead zone. 
Then, the wall is heated efficiently by stellar irradiation or the surface layers. As a result, 
the wall becomes thermally hot and leads to the back heating of the dead zone at smaller disk radii, 
which produces a positive temperature gradient in the dead zone.

We then use these results to examine how this thermal structure affects planetary migration. More specifically, 
we analytically calculate the timescale of type I migration, 
adopting the density and thermal structure of the disk with the dead zone. Figure  \ref{fig1} (panel (c); bottom) shows that 
the timescale of type I migration becomes {\it negative} in the dead zone where the positive temperature gradient is established - planets migrate 
{\it outward}. This occurs because such a temperature profile makes the inner resonances closer to a planet and the outer ones further away from the planet. 
As a result, the balance of the torque, that drives planetary migration, reverses there. Thus, when planets migrate either within or beyond dead zones, 
they will be trapped at the location where the net torque is zero. The location is often referred to as a planet trap (\cite[Masset et al 2006]{Masset06}).

\section{Unified Picture of Planet Traps}

As shown in this example, 
disk inhomogeneities play a significant role in planetary migration through the creation of planet traps. 
Since a number of disk inhomogeneities are already discussed in the literature, we take into account them and generalize how the disk inhomogeneities act as 
planet traps.

We adopt an analytical approach for comprehensively examining disk inhomogeneities and the resultant planet traps 
(\cite[Hasegawa \& Pudritz 2011]{HasegawaPudritz11}). We utilized analytical modeling of disk inhomogeneities and 
investigated how the direction of type I migration switches from inward to outward there. We consider three types of inhomogeneities: 
dead zones, ice lines, and heat transitions. We focus on water-ice lines, because the ice lines are most important 
for planet formation. Heat transitions are disk inhomogeneities at which the main heat source of protoplanetary disks changes from viscous heating to 
stellar irradiation. The temperature profile therefore changes from steep to shallow ones. 

We confirmed analytically that the net torque becomes zero at all the three types of disk inhomogeneities, 
so that they can act as planet traps for rapid type I migration. Also, we showed that the position of planet traps is sensitive to the surface density of disks, 
or equivalently the disk accretion rate. Figure \ref{fig1} (panel (d)) shows how the position of planet traps evolves with the disk accretion rate. 
It is important that single disks have up to three types of planet traps and that different traps move inward at different rates. 

Table \ref{table1} summarizes the position of planet traps for  various types of stars. The position of planet traps vary largely, 
depending on the mass of the central stars. Thus, our results indicate that stellar masses and disk accretion rates play a significant role in 
establishing a preferred (initial) scale of planetary systems.

\section{Implications of Planet Traps for Observations}

W have discussed how useful disk inhomogeneities and the resultant planet traps are for resolving the problem of rapid type I migration and
the origin of planetary system architectures. The next fundamental question is how such effects are tested against observations. In order to 
address such a question, we have recently computed evolutionary tracks of planets that grow at planet traps. We have found that, 
when planet traps are coupled with the core accretion scenario, 
the observations of exoplanets done by radial velocity techniques can be reproduced very well 
(\cite[Hasegawa \& Pudritz 2012]{HasegawaPudritz12}; and Pudritz \& Hasegawa, these Proceedings).  
As discussed above, the presence of dead zones alters the disk structure significantly, so that some kind of features may be shown in the observables. 
One possibility is in the SEDs/images of disks with inhomogeneities. We are currently investigating this for further supporting the importance of disk inhomogeneities and the resultant planet traps.

\begin{table}
%\begin{table*}
%\begin{minipage}{17cm}
\begin{center}
\caption{The position of planet traps. Adapted from \cite[Hasegawa \& Pudritz (2011)]{HasegawaPudritz11}}
\label{table1}
\begin{tabular}{cccc}
\hline 
                                     & Herbig Ae/Be stars                              \\
                                     & $r_{edge}$ (AU) & $r_{il}$ (AU) & $r_{ht}$ (AU) \\ \hline
$\dot{M}=10^{-4}$($M_{\odot}$/year) & 3.9             & 155           & 830           \\
$\dot{M}=10^{-6}$($M_{\odot}$/year) & 1.2             & 20            & 47            \\ 
\hline 
                                     & CTTSs                                           \\
                                     & $r_{edge}$ (AU) & $r_{il}$ (AU) & $r_{ht}$ (AU) \\ \hline
$\dot{M}=10^{-6}$($M_{\odot}$/year) & 3.7             & 11.7          & 42.4          \\
$\dot{M}=10^{-8}$($M_{\odot}$/year) & 1.2             & 1.5           & 2.4           \\
\hline 
                                     & M stars                                         \\
                                     & $r_{edge}$ (AU) & $r_{il}$ (AU) & $r_{ht}$ (AU) \\ \hline
$\dot{M}=10^{-8}$($M_{\odot}$/year) & 4.2             & N/A           & N/A           \\
$\dot{M}=10^{-9}$($M_{\odot}$/year) & 2.3             & 1.5           & N/A           \\
\hline 
\end{tabular} 

N/A denotes the case that planet traps excited by the disk inhomogeneities are not active due to convergence with a dead zone trap.
\end{center}
%\end{minipage}
%\end{table*}
\end{table}


\begin{thebibliography}{}

\bibitem[Gammie (1996)]{Gammie96}
{Gammie, C.F.} 1996,
\textit{ApJ}, 457, 355 

\bibitem[Hasegawa \& Pudritz (2010a)]{HasegawaPudritz10a}
{Hasegawa, Y., \& Pudritz, R.E.} 2010a,
\textit{ApJ}, 710, L167 

\bibitem[Hasegawa \& Pudritz (2010b)]{HasegawaPudritz10b}
{Hasegawa, Y., \& Pudritz, R.E.} 2010b,
\textit{MNRAS}, 401, 143 

\bibitem[Hasegawa \& Pudritz (2011)]{HasegawaPudritz11}
{Hasegawa, Y., \& Pudritz, R.E.} 2011,
\textit{MNRAS}, 417, 1236

\bibitem[Hasegawa \& Pudritz (2012)]{HasegawaPudritz12}
{Hasegawa, Y., \& Pudritz, R.E.} 2012,
\textit{ApJ}, 760, 117

\bibitem[Kley \& Nelson (2012)]{KleyNelson12}
{Kley, W., \& Nelson, R.P.} 2012,
\textit{ARAA}, 50, 211 

\bibitem[Masset et al (2006)]{Masset06}
{Masset, F.S. et al.} 2006,
\textit{ApJ}, 642, 478 

\bibitem[Qi et al (2011)]{Qi11}
{Qi, C. et al.} 2011,
\textit{ApJ}, 740, 84 

\bibitem[Tamura (2009)]{Tamura09}
{Tamura, M.} 2009,
\textit{Proceedings of American Institute of Physics Conference Series}, 1158, 11


\end{thebibliography}
\end{document}